# Ultra-Efficient Resistance Switching between Charge Ordered Phases in 1$T$-TaS$_2$ with a Single Picosecond Electrical Pulse

Rok Venturini,[1,2,*] Anže Mraz,[1,3] Igor Vaskivskyi,[1] Yevhenii Vaskivskyi,[1,2] Damjan Svetin,[1,4] Tomaž Mertelj,[1] Leon Pavlovič,[5] Jing Cheng,[6] Genyu Chen,[6] Priyanthi Amarasinghe,[7] Syed B. Qadri,[8] Sudhir B. Trivedi,[7] Roman Sobolewski,[6,9] and Dragan Mihailovic[1,4,*]

[1]*Jožef Stefan Institute, Jamova 39, Ljubljana, SI-1000, Slovenia,*

[2]*Faculty of Mathematics and Physics, University of Ljubljana, Jadranska 19, Ljubljana, SI-1000, Slovenia*

[3]*Faculty for Electrical Engineering, University of Ljubljana, Tržaška 25, SI-1000 Ljubljana, Slovenia*

[4]*CENN Nanocenter, Jamova 39, SI-1000 Ljubljana, SI-1000, Slovenia*

[5]*ELEP Electronics, Ljubljana, SI-1000, Slovenia*

[6]*Materials Science Program and Laboratory for Laser Energetics, University of Rochester, New York 14627, USA*

[7]*Brimrose Technology Corporation, Sparks, MD 21152, USA*

[8]*U.S. Naval Research Laboratory, Washington D.C., USA*

[9]*Department of Electrical and Computer Engineering and Department of Physics and Astronomy, University of Rochester, New York 14627, USA*

*Authors to whom correspondence should be addressed: rok.veturini@ijs.si and dragan.mihailovic@ijs.si

**Progress in high-performance computing demands significant advances in memory technology. Among novel memory technologies that promise efficient device operation on a sub-ns timescale, resistance switching between charge ordered phases of 1$T$-TaS$_2$ has shown to be potentially useful for development of high-speed, energy efficient non-volatile memory devices. Measurement of the electrical operation of such devices in the picosecond regime is technically challenging and hitherto still largely unexplored. Here we use an optoelectronic "laboratory-on-a-chip" experiment for measurements of ultrafast memory switching, enabling accurate measurement of *electrical* switching parameters with 100 fs temporal resolution. Photoexcitation and electro-optic sampling on a (Cd,Mn)Te substrate are used to generate and, subsequently, measure electrical pulse propagation, with intra-band excitation and sub-gap probing, respectively. We**

**demonstrate high contrast non-volatile resistance switching from high to low resistance states of a 1$T$-TaS$_2$ device using *single* sub-2ps electrical pulses. Using detailed modeling we find that the switching energy density per unit area is exceptionally small, $E_A = 9.4$ fJ/µm$^2$. The speed and energy efficiency of an electronic "write" process, place the 1$T$-TaS$_2$ devices into a category of their own among new generation non-volatile memory devices.**

Memory speed and energy efficiency have become one of the major bottlenecks limiting advanced computer development.[1] As a result, there has been significant interest in recent years to address this issue by exploring novel memory concepts, such as memristors,[2] phase-change,[3] ferroelectric[4] and magnetic memory[5] devices. One of the emerging new platforms for non-volatile storage of information is switching between distinct charge configurations of charge density wave (CDW) phases.[6–11] Such phases are common in low-dimensional materials and have attracted much interest from the point of view of fundamental physics as their formation is often not fully understood.[12,13] When a CDW order in a material is disrupted, for example by the hydrostatic pressure[14–16] or chemical doping,[17–19] novel material functionalities can emerge such as a large change in the electrical resistance or an appearance of the superconducting state. Optical, THz, or electrical pulsed excitations can also be used to drive the material towards a novel charge order.[6,7,20,21] However, the usefulness of such switching has long been limited, as excited CDW states are typically not stable, particularly at high temperatures.[9,22,23]

The material used in this study is 1$T$-TaS$_2$, a layered transitional metal dichalcogenide. At room temperature, the material is in the metallic nearly commensurate (NC) CDW state, from which the insulating commensurate (C) CDW state forms below around 180 K upon cooling. Upon heating the crystal from low temperature, the transition from the C state to the metallic stripe phase appears at around 220 K, followed by the transition to the NC state at around 280 K.[13,24] Either a laser or an electrical pulse excitation can drive the material from the insulating C state to the so-called metallic hidden (H) state with a complex pattern of domains separated by a network of domain walls.[6,7,11,25,26] The H state is metastable and is not observed in the equilibrium phase diagram of the material. Owing to topological protection, the H state is remarkably stable at low temperatures.[26] Relaxation back to the C state is temperature dependent, but the state is stable on laboratory timescales at temperatures lower than 20 K.[10,27,28] Robust memory device operation by switching between the two phases by electrical pulses of various lengths has already been demonstrated.[7,10,11] Optical experiments show that once the transition between the charge ordered phases is triggered, it can stabilize into the new order within 400 fs.[28] If picosecond *electrical* pulses could trigger the resistance-switching charge reconfiguration it may be possible to write to such memory devices at THz frequencies. Due to limitations of electrical generators and dispersion of cryostat cables, demonstration of electronic device operation on a such short timescale is currently still very challenging.

A way to overcome these issues is by *in situ* generation and detection of ultrashort electrical pulses on-a-chip by femtosecond laser pulses and then propagating electrical pulses along a coplanar transmission line (TL) to the device. While this technology is known[29–32] and has been used to perform switching of superconducting Josephson junctions[33] and circuits,[34,35] interest in this topic has recently re-emerged with ultrafast electronic studies of topological states of matter[36,37] and magnetization reversal.[38,39] In our case, we embedded a $1T$-TaS$_2$ device in the TL on a substrate that is photoconductive and simultaneously exhibits an electro-optic (EO) effect; therefore in addition to the DC resistance measurement, ultrashort electrical pulses can be generated and sampled anywhere along the TL. The EO detection can easily be calibrated to a known voltage, so the measured picosecond pulse voltage can be used to calculate the pulse energy. This offers a significant advantage over experiments that use either a photoconductive gap or an antenna for detection, where calibration is difficult and the energy for switching can only be given as the upper limit estimate.[38,39]

To facilitate both electrical pulse generation and EO detection on the same substrate, we used a vanadium-doped Cd$_{0.55}$Mn$_{0.45}$Te (CMT) single crystal substrate grown by the vertical Bridgman method.[40] CMT is a semiconductor with a 2.02 eV wide band gap as determined by the transmittance measurement [Fig. 1(a)]. Figure 1 (b) shows a schematic of the laboratory-on-a-chip TL circuit used in our experiment. First, a 32 nm thick flake is exfoliated from a bulk $1T$-TaS$_2$ crystal and deposited on the substrate. Next, the TL circuit is lithographically fabricated on top by deposition of 3 nm of Cr followed by 160 nm of Au. The fabricated electrodes are marked A, B and C, where A is the ground, B is used to provide a DC voltage, and C is used to measure the resistance of the $1T$-TaS$_2$. TL electrodes A and B are 12 µm wide and 34 µm apart. The $1T$-TaS$_2$ flake is positioned 100 µm from the C electrode, with the TL gap of 0.6 µm above the flake. The crucial part of the circuit is a 22 µm wide gap between electrodes B and C that acts as a capacitor, which prevents the DC flow through the $1T$-TaS$_2$ flake when a DC bias voltage is applied between leads A and B. At the same time, it allows the short photogenerated electrical pulses to be transmitted with low attenuation. The entire TL is 4-mm-long (not shown) to reduce the impact of pulse reflections from the ends.

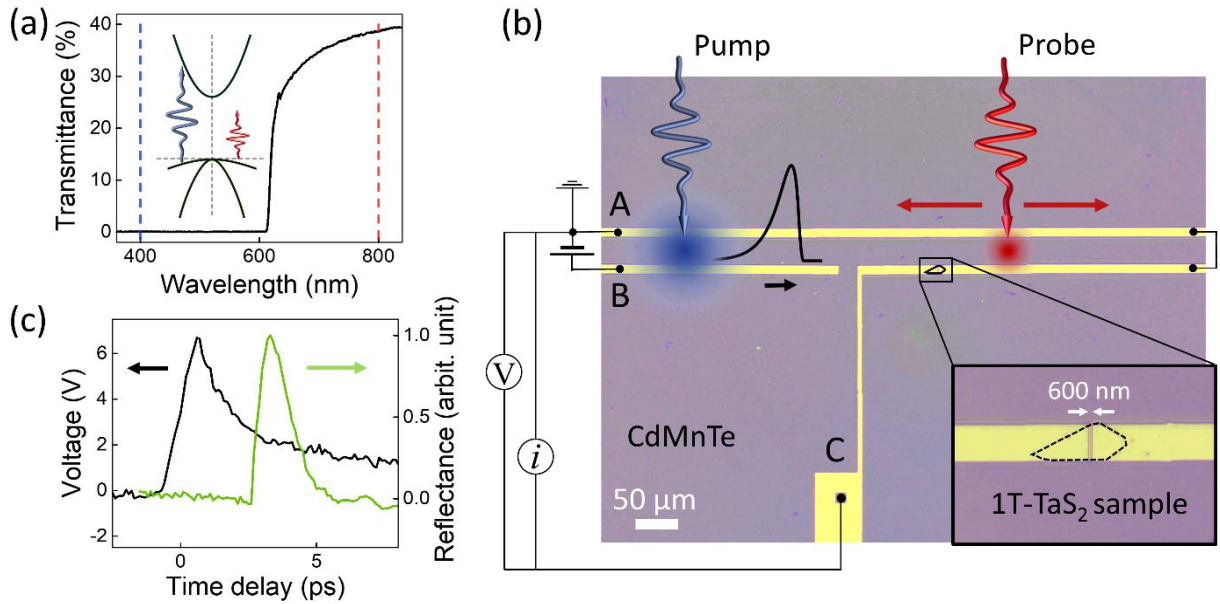

**FIG. 1.** (a) The absorption spectrum of the CMT substrate in relation to the pump beam (blue) and the probe beam (red). An inset shows a schematic of the CMT band structure around k = 0.[41] (b) A microscope image of the fabricated TL circuit over the exfoliated 1$T$-TaS$_2$ flake with the pump beam (blue) and the span of the probe beam (red) positions. The voltmeter and the current source are used for measurement of the flake resistance. The voltage source provides a DC bias voltage for electrical pulse generation. The inset shows a zoomed-in view of the 1$T$-TaS$_2$ flake in the circuit. (c) The time-resolved transient reflectivity measurement (green) of the CMT substrate away from the TL circuit by spatially overlapping pump and probe beams. The time-resolved measurement of a picosecond electrical pulse (black) detected by the probe beam 40 µm in front of the 1$T$-TaS$_2$.

The area between electrodes A and B that is used as a photoswitch [Fig. 1(b)] shows ohmic behavior without laser illumination with a resistance greater than $10^9$ Ω. An electrical pulse is excited by a 400 nm (3.1 eV) laser beam (pump) excitation of the photoswitch. Ultrashort carrier lifetime of a substrate is needed for creation of picosecond long electrical pulses. Carrier lifetime measurement of the CMT substrate by the (400 nm/800 nm) pump-probe reflectivity measurement away from the TL circuit [Fig. 1(c)] reveals sub-2-ps carrier lifetime. In addition, the CMT substrate has an EO tensor component along the [001] direction so a propagating electric field along the TL fabricated on the (110) plane modifies the substrate birefringence via the Pockels effect.[32] The change in birefringence is measured via the change in polarization of an 800 nm (1.55 eV) laser beam (probe) in transmission. Both the excitation of the picosecond electrical pulses by the pump beam and the EO sampling by the probe are done with ~100 fs long laser pulses at a 250 kHz repetition rate. An acousto-optic modulator positioned in the laser beam allows for single pulse picking, when necessary. The temporal evolution of the electrical pulses is measured by adjusting the time delay between the pump and probe pulses. The EO response is calibrated to a voltage between electrodes by measuring the EO response to a continuous-wave 27 kHz sine signal with the amplitude of 0.25 V.

We first present room-temperature characterization of the picosecond electrical pulses generated in the TL circuit. Using a low laser fluence of 0.3 mJ/cm$^2$ and a moderate DC voltage of 76 V between the electrodes, we focus the probe beam 40 µm from the 1$T$-TaS$_2$ flake to give an accurate measurement of the electrical pulse shape propagating across the device. In this configuration, the full width at half maximum (FWHM) of the electrical pulse is measured to be 1.9 ps as shown in Fig. 1(c)–black trace.

For ultrafast resistance switching of the 1$T$-TaS$_2$ device, the chip is cooled from room temperature to 16 K, while the DC resistance of the device is continuously measured (Fig. 2). At 16 K we position the pump beam between the TL electrodes A and B, as before, and the probe after the 1$T$-TaS$_2$ crystal. We measure the propagating electrical pulse shape with low DC bias voltage to verify that the pulse shape is unchanged. We then increase the DC voltage bias to 220 V without pump illumination. Using a single optical pump pulse with the fluence of 5.5 mJ/cm$^2$ from an acousto-optic pulse-picker, we now excite a single electrical switching pulse. After exposure we observe a large drop in the device resistance as shown in Fig. 2 (black arrow). This is consistent with resistance switching observed previously using longer electrical pulses.[7,8] During heating of the crystal back to room temperature we observe an increase in the sample resistance between 50 and 100 K corresponding to the relaxation of the metastable H state back to the C state.[6,27] The cycle of cooling–switching–erasing was sequentially repeated many times, demonstrating consistency of single-shot ultrafast resistance switching. Faster erase sequences for effective device operation can be done with another electrical pulse and have been discussed in previous work. [8,10]

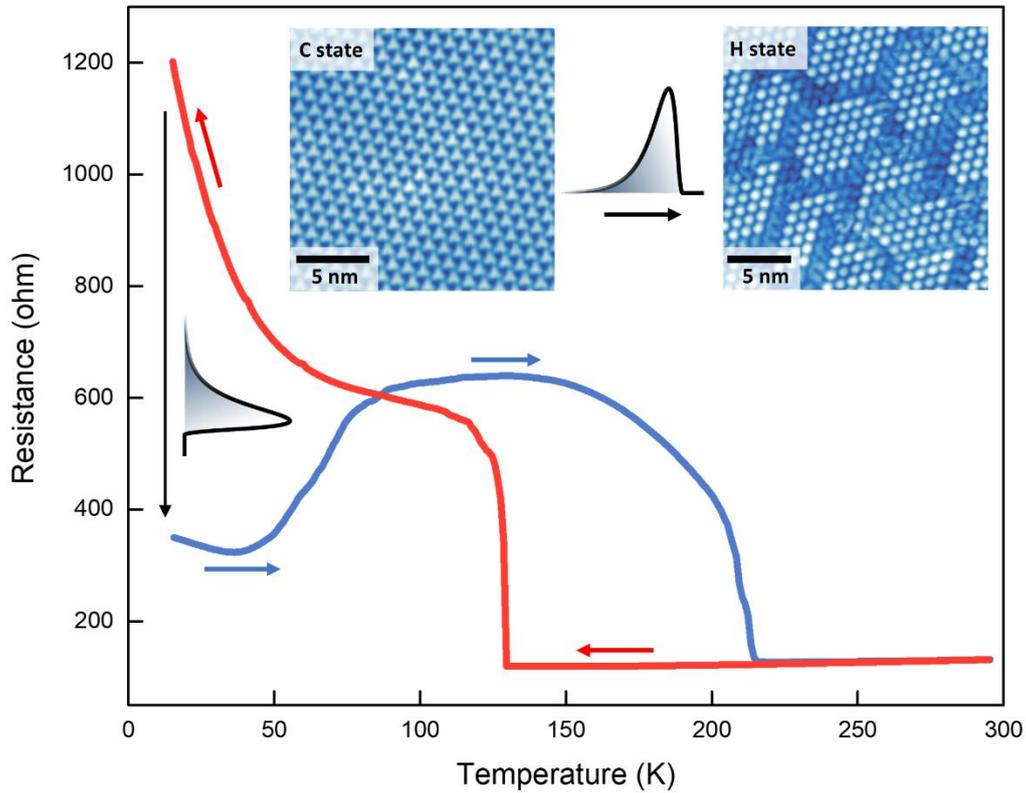

**FIG. 2.** Resistance of the 1$T$-TaS$_2$ flake upon cooling (red), resistance switching of the flake by a single 1.9 ps electrical pulse at 16 K (black arrow) and heating of the flake back to room temperature (blue) results in H state relaxation back to the C state between temperatures of 50 K and 100 K. The insets are scanning tunneling microscope images of the C state (left) and the H state (right) presented for illustration of charge reordering in the switching process.[42]

In all further switching experiments the temperature was kept below 20 K with the 1$T$-TaS$_2$ device operation in a non-volatile mode. Next, we investigate how the resistance drop depends on the picosecond electrical pulse amplitude. We vary the picosecond pulse amplitude by either changing the DC bias voltage or the pump fluence. The pulse amplitude (at a fixed pump fluence of 0.4 mJ/cm$^2$) increases almost linearly with increasing the DC voltage as shown in Fig. 3(a) (blue), matching previous predictions and observations.[43] By varying the pump fluence (at a fixed DC bias voltage of 60 V) a sublinear dependence is observed, as shown in Fig. 3(b) (blue), which is likely due to a saturation of absorption of the pump photons and is consistent with previous results for the CMT substrate.[32] Furthermore, no change in the shape of the picosecond pulse is observed by changing the DC bias and/or the pump fluence within the explored intervals of 0-250 V and 0-5.5 mJ/cm$^2$, respectively. In both cases we observe that the flake resistance remains unchanged for the picosecond pulse amplitudes below a DC bias threshold of ~70 V while above the threshold, it continuously decreases with increasing electrical pulse amplitude, as shown in Fig. 3(a) and (b).

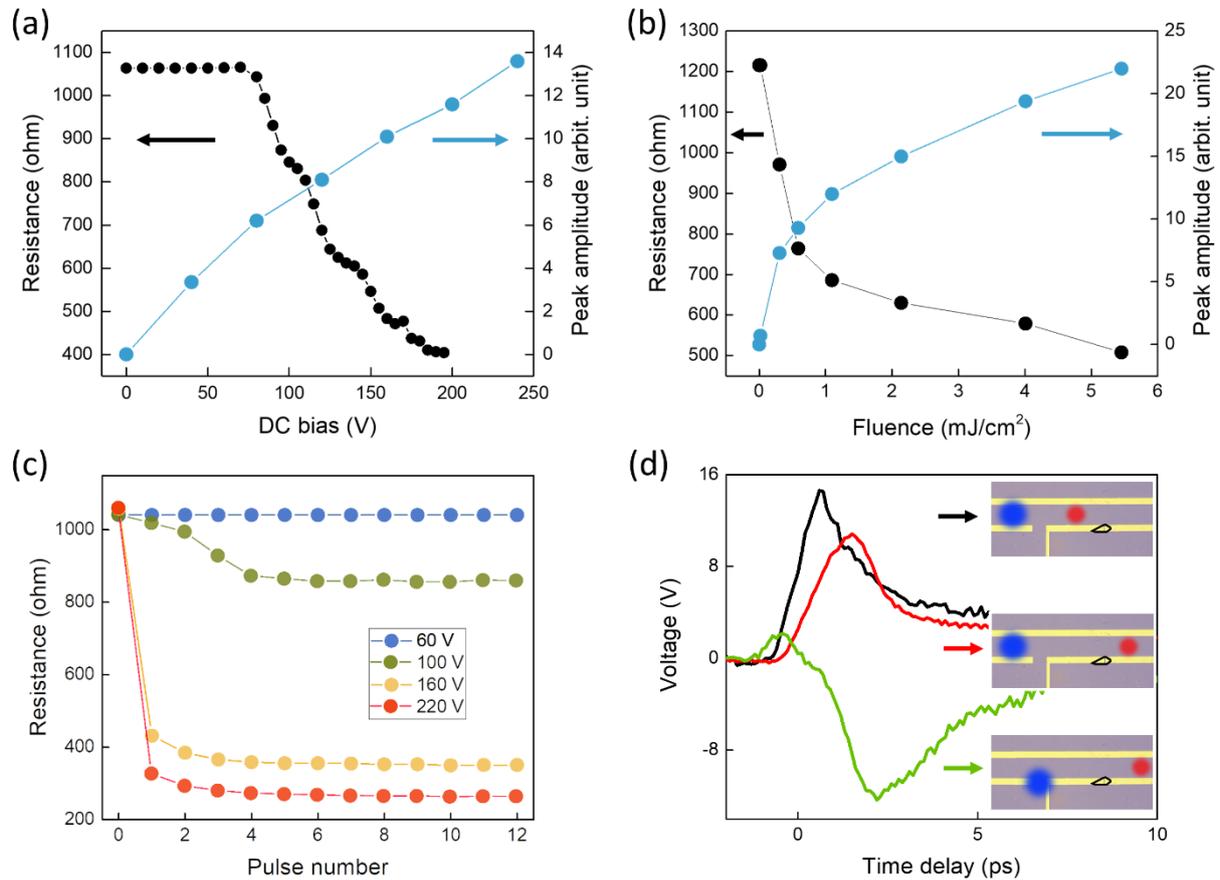

**FIG 3.** (a) The electrical resistance of the 1*T*-TaS$_2$ flake (black) and picosecond pulse amplitude (blue) in relation to the DC bias at the constant pump fluence. (b) The electrical resistance of the flake (black) and excited picosecond pulse amplitude (blue) in relation to the pump fluence at the constant DC bias. (c) Build-up effect on the 1*T*-TaS$_2$ resistance by applying multiple electrical pulses one by one at different DC bias voltages. (d) The shape of the picosecond electrical pulse sampled by the probe beam 40 µm before (black) and 100 µm after the sample (red). The pump beam on the gap excites the electrical pulse with reversed polarity (green). Inset figures show the optical microscope image of the TL with the schematic of the pump (blue) and probe (red) beam positions.

Furthermore, we investigate the 1*T*-TaS$_2$ flake resistance changes in response to repeated single-pulse excitation at different DC bias voltages. After each single-pulse exposure we measure the resistance and wait 1 min before applying the next pulse. As shown in Fig. 3(c), the largest resistance drop is observed after the initial pulse, with the next few pulses gradually lowering the resistance toward saturation. Below the threshold, we do not observe any change in the flake resistance irrespective of the number of the applied pulses. Even after applying $10^8$ pulses at a 250 kHz repetition rate below the threshold, at 60 V DC bias, there was no detectable resistance change of the device. This demonstrates that the threshold for switching onset is very sharp and the device behavior is robust even for such short pulses.

Finally, we perform a slightly modified electrical switching experiment where instead of focusing the pump beam between electrodes A and B we focus it in the gap between electrodes B and C as shown in Fig. 3(d). We observe a similar pulse shape with the reversed polarity. Repeating the switching experiment with the electrical pulses of the reversed polarity we obtain a very similar switching behavior.

Repeating the experiment with the probe beam on both sides of the 1$T$-TaS$_2$ crystal, we observe that the pulse shape is unaltered [Fig. 3(d)] as it travels past the flake, but the amplitude is slightly smaller due to high-frequency losses.[32] The latter means, that only a small fraction of the picosecond pulse energy is absorbed in the 1$T$-TaS$_2$ flake while the majority of the pulse propagates along the TL past the flake. A large random spot-to-spot variation of the EO signal (~30%) prevents us from accurately measuring the very small dissipated energy in the 1$T$-TaS$_2$ flake by directly comparing the incident and transmitted EO signal energies. Instead, to estimate the energy of switching of the 1$T$-TaS$_2$ flake, we employ 3D electromagnetic simulation software (Ansys HFSS). First, we simulate a straight coplanar TL without any gaps and the flake. For the CMT substrate we take the dielectric constant obtained in the parent compound.[44] From the simulation we obtain the TL impedance of 125 Ω and the pulse propagation velocity of $c = 1.25 \times 10^8$ m/s. The calculated velocity is close to the measured one of $1.2 \times 10^8$ m/s. Using the calculated impedance and the experimentally measured pulse voltage between the electrodes we calculate the energy of the propagating picosecond electrical pulse at the threshold for switching, $E_T = \int \frac{U_{TL}^2(t)}{Z} dt = 0.59$ pJ, where $U_{TL}$ is the voltage between the TL electrodes and Z is the impedance of the TL. The integral is extended to 7 ps after the pulse start to include also the pulse tail.

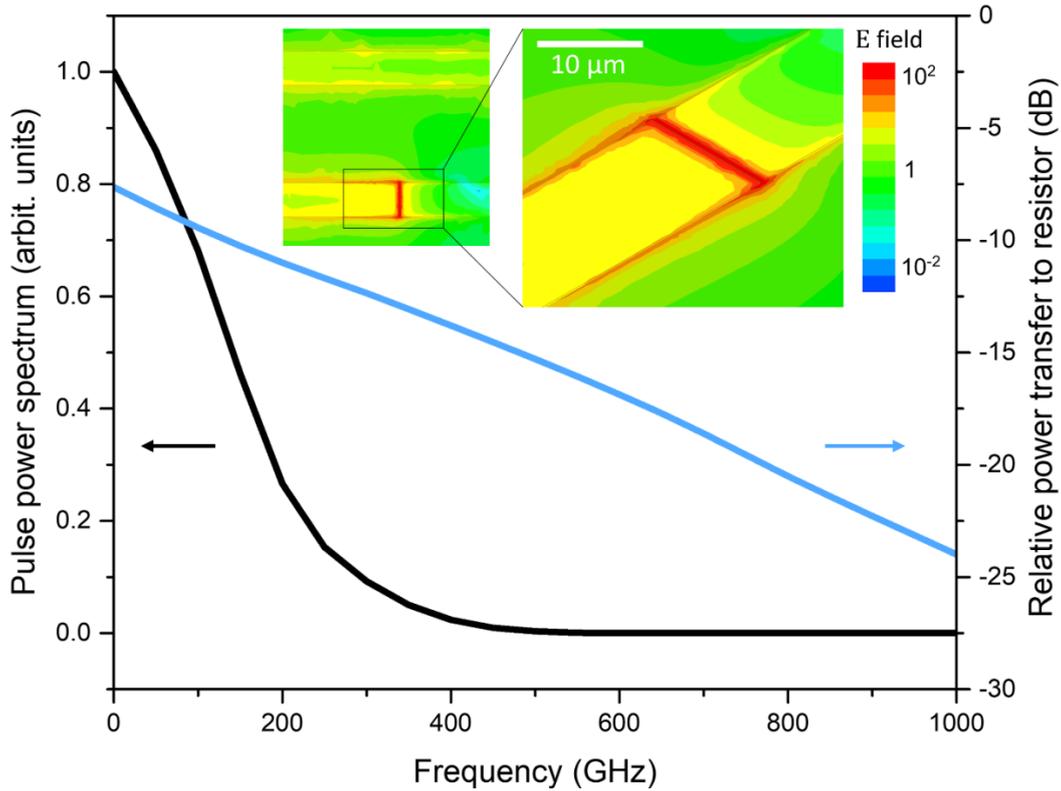

**FIG. 4.** The power spectrum of the picosecond electrical pulse (black) and the relative power transfer of the propagating pulse to the 1$T$-TaS$_2$ flake obtained from the simulation (blue). The insets show the calculated electric field relative to the field between TL electrodes simulated at the 100 GHz frequency.

Next, to estimate the relative absorbed energy, we modify the TL in the simulation by inserting a 600 nm gap using a 1100 Ω 2D resistor to represent the 1$T$-TaS$_2$ flake in the high resistance state (see inset in Fig. 4). The simulated relative power transfer to the resistor inside the gap is found to be small and strongly frequency dependent (see Fig. 4). The experimental dissipated energy is then estimated by multiplying the experimental picosecond electrical pulse power spectrum (Fig. 4) by the simulated power-dissipation spectrum in the resistor and integrated over the frequency. We find that only 11.5 % of the electrical pulse energy is dissipated in the 1$T$-TaS$_2$ flake. For the electrical pulse amplitude at the switching threshold, this corresponds to switching energy of 68 fJ. Note that this is an upper bound since it assumes that all the energy is dissipated in the switching process. Optical experiments indicate that the transition to a low resistance state is triggered already within 400 fs after the optical pulse arrival, which indicates that only the first part of the electrical pulse might be responsible for switching, and the energy from the rest of the electrical pulse only contributes to heating of the already switched device.

Using the dimensions of the 1$T$-TaS$_2$ flake between the electrodes, the estimated switching energy of 68 fJ corresponds to an energy density per unit volume $E_V = 0.3$ pJ/µm$^3$, or an energy per unit area $E_A = 9.4$ fJ/µm$^2$. Thus, it appears that 1$T$-TaS$_2$ devices exhibit the lowest switching energy per unit area and the fastest 'write' time for any memory device reported.[10] Considering the large size of the device used in the present experiments, the energy for resistance switching could potentially be further reduced by more than an order of magnitude by fabricating smaller devices.

To conclude, our laboratory-on-a-chip experiments demonstrate that switching to a non-volatile state with a large resistance contrast is possible in 1$T$-TaS$_2$ devices approaching terahertz speeds, presently limited only by the substrate carrier lifetime. Direct, high temporal resolution measurements of the pulse propagation on the transmission line circuit are the key to demonstrating ultrafast and energy efficient switching from a high to low resistance state, approaching the optical 400 fs switching limit.[28] Based on the demonstration of optical erase with 50 ps pulses,[6] and the demonstration of fast electrical erase via Joule heating[8,10] there are no apparent fundamental limitations to achieving fast operation in both directions.

Considering that the devices operate at low temperatures, the charge configuration memory elements based on the resistance switching in 1$T$-TaS$_2$ could fit well with the emerging technologies that operate at cryogenic temperatures and lack a suitable memory device.[45] Due to ultra-high switching speed, one obvious match is the superconducting single-flux-quantum (SFQ) logic, which relies on picosecond long SFQ pulses to process data.[45,46] With the ability to grow few-layer crystals by epitaxial growth,[47,48] devices based on the charge configuration switching could provide a much-needed ultrafast and energy efficient solution for classical information storage in both the classical and quantum computing environments.

**Conflict of interest:** The authors declare no competing interests.

**Acknowledgment**

The authors would like to thank Jan Ravnik for his help in building the optical setup. The work at the Jožef Stefan Institute was supported by the Slovenian Research Agency (No. P1-0040, No. N1-0092, No. J1-2455, R.V. to No. PR-10496, A.M. to No. PR-08972, D.S. to No. I0-0005), Slovene Ministry of Science (No. Raziskovalci-2.1-IJS-952005), and ERC PoC (No.GA767176). The work at Rochester and Brimrose was supported by the DOE STTR 2021 Phase 1 Grant No. DE-SC0021468. We thank the CENN Nanocenter for the use of the AFM and the FIB.